\documentclass[conference]{IEEEtran}
\usepackage[T1]{fontenc}
\usepackage{cite}
\usepackage{graphicx}
\usepackage{amsmath,amssymb}
\usepackage{algorithm}
\usepackage{algpseudocode}
\usepackage{hyperref}
\usepackage{booktabs}
\usepackage{multirow}
\begin{document}

\title{Detection-Guided Adaptive Purification with Diffusion Models for Robust Audio Deepfake Detection}

\author{
\IEEEauthorblockN{Muhammed Salih Kayhan$^{1}$ and Qiben Yan$^{1}$}
\IEEEauthorblockA{$^{1}$Michigan State University, East Lansing, MI, USA\\
\{kayhanmu,qyan\}@msu.edu}
}

\maketitle

\begin{abstract}
Audio deepfake detectors remain vulnerable to adversarial perturbations that suppress the acoustic cues used for detection, allowing manipulated utterances to evade the detector. Although existing defenses can improve robustness, they require retraining the detector or introduce additional distortion. Diffusion-based purification instead leaves the pretrained detector unchanged, but existing methods use the same purification strength for all inputs, creating a trade-off between removing adversarial perturbations and preserving the subtle spoofing cues needed for detection. In this paper, we propose Detection-Guided Adaptive Purification (DGAP), a diffusion-based defense that adjusts purification strength per input. Building on the observation that a light purification perturbs the detector score of an adversarial input far more than that of a benign one, the framework uses the resulting score shift as a reference-free indicator of adversarial manipulation. Inputs with small shifts are passed unchanged, whereas flagged inputs undergo stronger purification before final detection. We evaluate the framework against three adversarial attack settings across three deepfake detectors, and compare it with nine existing defenses. Our results show that DGAP achieves the best defense performance across all detectors while leaving benign inputs nearly unaffected, and remains effective under the defense-aware adaptive attack. 

\end{abstract}


\section{Introduction}
Speech is widely used in digital communication, biometric identity verification, and voice-enabled services, where the authenticity of an utterance can directly affect trust and security. At the same time, recent advances in text-to-speech (TTS), voice conversion (VC), voice cloning systems, and speech foundation models have made it possible to generate highly realistic audio deepfakes that closely imitate real speakers. Such deepfakes can falsely represent a person's identity, manipulate spoken
evidence, or enable fraudulent decision-making in sensitive
services~\cite{farooq2024securing,diel2025harm}. As publicly available generation tools lower the barrier to creating convincing fake speech, reliable audio deepfake detection becomes increasingly critical under both benign and adversarial conditions.

\begin{figure}[t]
    \centering
    \includegraphics[width=\linewidth]{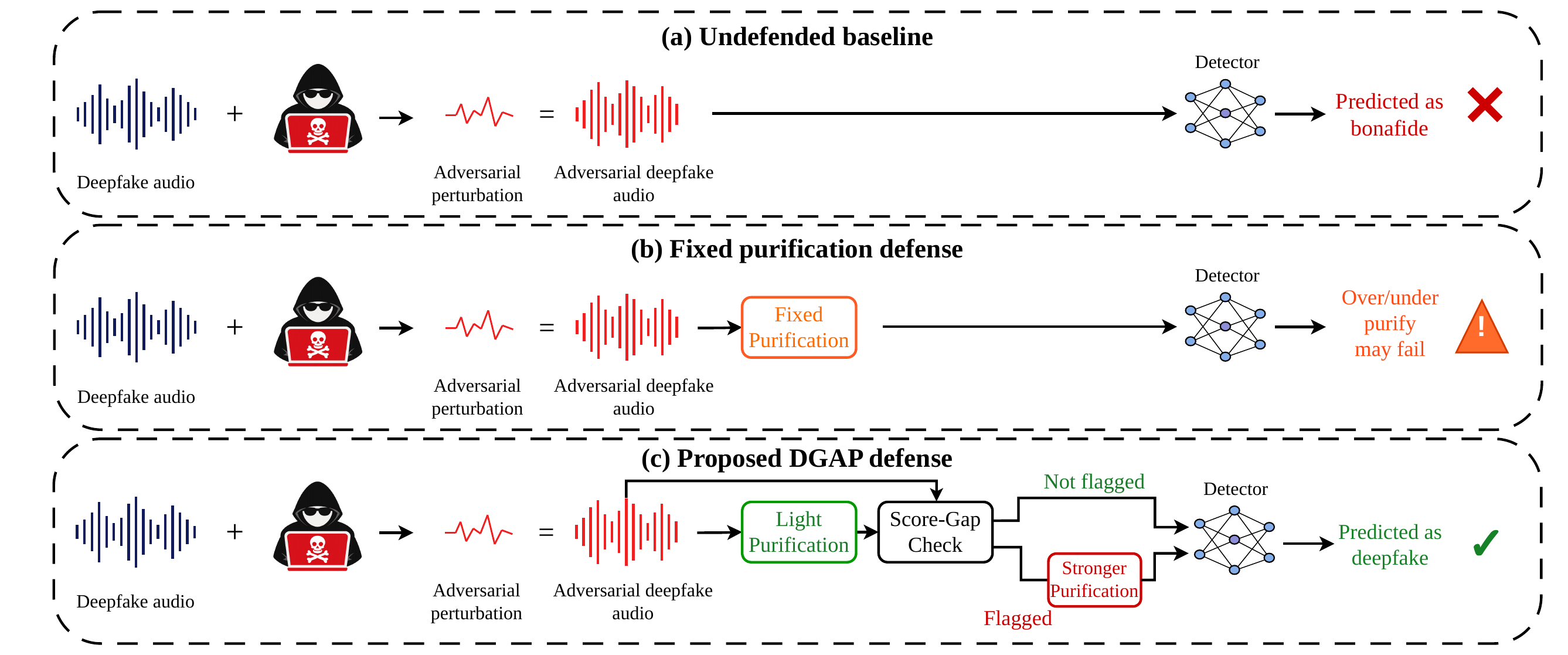}
    \caption{Comparison of (a) an undefended baseline, (b) a fixed purification defense, and (c) the proposed DGAP defense under adversarial deepfake attacks.}
    \label{fig:overview}
\end{figure}

A wide range of audio deepfake detection methods have been proposed, most relying on deep neural networks (DNN) and acoustic or learned representations~\cite{hua2021towards,tak2021end,wang2021comparative}, and evaluated on benchmarks such as
ASVspoof2019~\cite{todisco2019asvspoof} and
In-the-Wild~\cite{muller2022does}. Their reliability, however, remains uncertain when inputs are deliberately crafted to evade detection. An attacker does not need to generate a new deepfake from scratch, since small and often imperceptible perturbations added to an already manipulated utterance can suppress or distort the acoustic cues these methods rely on, causing a deepfake that would normally be detected to be misclassified as bonafide speech, as illustrated in Fig.~\ref{fig:overview}(a). This exposes a gap between benchmark-level accuracy and the robustness required for secure deployment.

Robustness against such perturbations has been studied across image and audio domains, with defenses including input-level
transformations~\cite{guo2017countering,xie2017mitigating} and
signal-level preprocessing and noise injection~\cite{bai2021recent}. In audio deepfake detection specifically, countermeasures such as adversarial training~\cite{kawa2022defense} and inference-time noise
addition~\cite{kim2025inference} have been explored. These approaches can improve robustness in specific settings, but they may require attack-aware training, introduce additional signal distortion, or rely on assumptions about the attack and deployment scenario.

Recently, input purification with diffusion models has emerged as a promising approach for defending audio models without
retraining~\cite{wu2023defending,guo2024wavepurifier}. Existing
diffusion-based defenses, however, typically apply a fixed
purification strength to all inputs (Fig.~\ref{fig:overview}(b)). This is not ideal for audio deepfake detection, where excessive purification may also remove the subtle artifacts that indicate synthetic speech, whereas weak purification may preserve these artifacts but fail to remove the adversarial perturbation. This trade-off motivates an adaptive strategy that adjusts the purification strength according to the input rather than treating all inputs uniformly.

In this paper, we propose an adaptive
purification framework, called detection-guided adaptive purification (DGAP), as shown in Fig.~\ref{fig:overview}(c), that
uses a diffusion model to purify incoming audio before it reaches the
deepfake detector. Our approach first applies a light purification
step and observes how the detector's response changes. If the score
shifts significantly, the input is flagged and subjected to stronger
purification, while the remaining inputs pass unchanged. In this way,
the framework concentrates the purification effort where it is needed,
neutralizing adversarial perturbations while preserving the cues the
detector relies on to identify fake speech.

Our main contributions are summarized as follows:
\begin{itemize}
\item \textbf{Adaptive Purification Framework:} We develop DGAP, a
detection-guided adaptive purification framework for robust audio
deepfake detection under adversarial perturbations. The framework
purifies each input with a diffusion model at a strength selected from the detector's own response, without modifying or retraining the detector.
\item \textbf{Gating Mechanism:} We introduce a reference-free gating mechanism that flags adversarially perturbed inputs before final detection. By comparing the detector scores of the original audio and its lightly purified version, the gate requires no clean reference and no knowledge of the attack.
\item \textbf{Extensive Evaluation:} We evaluate our framework against three white-box attack settings across three deepfake detectors, and compare it with nine existing defenses. Our framework achieves the strongest overall defense performance on all detectors while leaving benign inputs nearly unaffected, and remains effective under a defense-aware adaptive attack.
\end{itemize}
The source code is available at \href{https://github.com/msalih98/DGAP}{GitHub Repository}.

\section{Related Work}

\subsection{Audio Deepfake Detection}
Audio deepfake detection is typically formulated as a binary classification problem in which a detector determines whether an input utterance corresponds to genuine human speech, often referred to as bonafide speech, or spoofed speech generated by text-to-speech, voice conversion, or voice cloning. Early systems addressed this task with a modular pipeline, where an acoustic front-end first extracts descriptive features and a separate back-end classifier makes the authenticity decision. The front-end commonly represents the speech signal through spectral or cepstral features, including mel-frequency cepstral coefficients (MFCC)~\cite{chen2010speaker}, linear-frequency cepstral coefficients (LFCC)~\cite{tian2016spoofing}, constant-Q cepstral coefficients (CQCC)~\cite{todisco2017constant}, and constant-Q transform (CQT) features~\cite{cheng2019replay}. These representations are then passed to back-end classifiers such as Gaussian mixture models (GMM)~\cite{reynolds2000speaker} and support vector machines (SVM)~\cite{hearst1998support}. As the field progressed, stronger neural back-ends, including convolutional networks, LCNN~\cite{wu2020light}, ResNet~\cite{he2016deep}, Res2Net~\cite{gao2019res2net}, and DARTS~\cite{liu2018darts} based architectures~\cite{ge2021partially}, were introduced into pipeline-based systems. While such approaches offer a clear mechanism for capturing artifacts introduced by speech synthesis, their effectiveness depends strongly on how well the selected representation exposes those artifacts. Since deepfake cues can be weak, localized, or spread across time and frequency, a fixed front-end may not reveal all information required for reliable detection.

This dependency on pre-defined representations encouraged a move toward deep learning and end-to-end detection models that learn discriminative cues directly from raw or transformed audio. RawNet2, introduced by Jung et al.~\cite{jung2020improved}, processes raw waveforms using convolutional and recurrent layers, and Tak et al.~\cite{tak2021end} later adapted this architecture to anti-spoofing, demonstrating that competitive detection can be achieved without manually designed acoustic features. Graph-based detectors extended this direction by explicitly modeling relationships among cues appearing in different spectral sub-bands and temporal regions.  Specifically, RawGAT-ST~\cite{tak2021end1} captures temporal and spectral dependencies through parallel graph structures, and AASIST~\cite{jung2022aasist} combines heterogeneous spectro-temporal graph representations within an attention-based framework. Beyond these models, Khan et al.~\cite{khan2023spotnet} explored compact CNN architectures on mel-spectrograms for efficient deployment. Extending this line of work, they later introduced attention-based mechanisms~\cite{khan2024frame} for identifying frame-level inconsistencies. Consequently, self-supervised speech representations, including Wav2Vec~2.0~\cite{baevski2020wav2vec}, WavLM~\cite{chen2022wavlm}, and XLS-R~\cite{babu2021xls}, have also been incorporated as front-ends or embeddings, allowing detectors to exploit larger-scale pretraining and richer acoustic representations.

Despite these advances, strong benchmark performance does not necessarily imply robustness under adversarial conditions. Rabhi et al.~\cite{rabhi2024audio} showed that small and often imperceptible perturbations can mislead audio deepfake detectors and cause spoofed speech to be classified as bonafide. Farooq et al.~\cite{farooq2025transferable} further demonstrated that such perturbations can transfer across different audio deepfake detection models, suggesting that the vulnerability is not limited to any single detector architecture. These findings motivate the need for defense mechanisms designed specifically for adversarially manipulated deepfake audio.

\subsection{Adversarial Defenses and Purification}
The vulnerability in deep learning models to adversarial perturbations has motivated a wide range of defense strategies. A common approach is adversarial training, which incorporates adversarial examples into optimization to improve robustness. Madry et al.~\cite{madry2017towards} formulated this as robust optimization against worst-case perturbations, and later work extended this idea across attacks, architectures, and training
strategies~\cite{bai2021recent}. In audio deepfake detection, Kawa et al.~\cite{kawa2022defense} trained detectors on perturbed spoofed speech to improve adversarial resistance. However, retraining-based defenses are often tied to the attacks seen during
training~\cite{cheng2024stability} and become impractical when
protecting fixed, pretrained detectors against changing attack
settings.

Another line of work detects adversarial inputs before classification.
Metzen et al.~\cite{metzen2017detecting} introduced an auxiliary
network for recognizing adversarial perturbations, while Lee et
al.~\cite{lee2018simple} characterized adversarial and
out-of-distribution inputs through feature-space statistics. Related
ideas have also been explored in speech and audio deepfake settings,
including adversarial-sample detection for speaker
verification~\cite{li2020investigating} and auxiliary strategies for
audio deepfake robustness~\cite{uddin2025sheild}. These methods can
improve attack awareness, but they primarily decide whether an input
is adversarial rather than producing a corrected signal for the
downstream detector.

Purification-based defenses instead modify the input before
classification. Early work used fixed transformations such as
compression and smoothing~\cite{guo2017countering} and
quantization~\cite{lin2019defensive} to weaken imperceptible
perturbations. Similar input-level strategies have been explored in
speech, including denoising~\cite{chang2021defending},
high-frequency suppression~\cite{olivier2021high},
self-supervised re-synthesis~\cite{wu2021improving}, spatial
smoothing~\cite{wu2020defense}, and inference-time noise addition~\cite{kim2025inference}. Randomized smoothing similarly
classifies an input by averaging predictions over noisy
copies~\cite{cohen2019certified}. These defenses are attractive for
fixed detectors because they can be applied at test time, but their
effectiveness depends on a processing strength chosen in advance. Weak
processing may leave the adversarial perturbation effective, whereas
overly strong processing may distort the acoustic evidence needed to
distinguish spoofed from bonafide speech.

Diffusion-based purification replaces static transformations with a
learned reconstruction process. In the image domain,
DiffPure~\cite{nie2022diffusion} adds a controlled amount of noise to
an input and then reverses the diffusion process to remove adversarial
perturbations. In audio, diffusion-based purification has mainly been
studied for automatic speech recognition (ASR) rather than deepfake detection.
AudioPure~\cite{wu2023defending} considers single-word speech commands
in waveform and spectrogram domains, while
WavePurifier~\cite{guo2024wavepurifier} extends diffusion purification
to longer utterances through a hierarchical process. This transfer is
not straightforward for audio deepfake detection because ASR-oriented
purification aims to recover intelligible speech content, whereas
deepfake detectors rely on subtle synthesis-related cues. A
purification process that improves recognition can therefore remove
evidence needed for spoof detection.

Although diffusion models have recently been explored for audio
deepfake detection through generalization-oriented
reconstruction~\cite{cheng2026diffusion} and artifact
explanation~\cite{grinberg2025data}, their role as adaptive purifiers against adversarially perturbed deepfake audio remains underexplored. Existing diffusion-based defenses typically apply the same purification schedule to every input rather than adapting it to the detector's response. These observations motivate a detector-guided adaptive purification strategy that varies its strength per input for audio deepfake detection.

\section{Background}
\subsection{Diffusion Models}
\label{sec:diffusion}

Diffusion models~\cite{ho2020denoising,song2020score} are generative models that synthesize data by reversing a gradual noising process. A fixed forward process progressively transforms a clean sample into Gaussian noise, while a learned reverse process reconstructs it from noise. We denote a clean sample by $x_0\in\mathbb{R}^{d}$ drawn from the data distribution $p_0$, and let $p_t$ be the marginal distribution of its diffused version $x_t$ at level $t$.

\noindent\textbf{Forward process.}
The continuous-time diffusion models use a forward stochastic differential equation (SDE) on the horizon $[0,1]$ as follows~\cite{song2020score}: 
\begin{equation}
  \mathrm{d}x = f(x,t)\,\mathrm{d}t + g(t)\,\mathrm{d}w,
  \qquad t\in[0,1],
  \label{eq:sde}
\end{equation}
where $f$ is the drift coefficient, $g$ is the diffusion coefficient, and $w$ is a standard Wiener process. We adopt the variance-preserving SDE (VP-SDE), for which $f(x,t)=-\tfrac{1}{2}\beta(t)\,x$ and $g(t)=\sqrt{\beta(t)}$, with a linear noise schedule $\beta(t)=\beta_{\min}+(\beta_{\max}-\beta_{\min})t$. The forward process then admits the closed-form perturbation kernel: 
\begin{equation}
  x_t = \sqrt{\bar\alpha_t}\,x_0 + \sqrt{1-\bar\alpha_t}\,\epsilon,
  \quad \epsilon\sim\mathcal{N}(0,I),
  \quad \bar\alpha_t = e^{-\int_0^t \beta(u)\,\mathrm{d}u},
  \label{eq:kernel}
\end{equation}
which yields the noised sample at any level $t$ without integrating the
intervening dynamics. The signal-retention coefficient $\bar\alpha_t\in(0,1]$
decreases monotonically with $t$ and thus indexes the noise level, from near-complete retention ($\bar\alpha_t\!\to\!1$) to near-Gaussian noise ($\bar\alpha_t\!\to\!0$).%
\footnote{In practice, the SDE is discretized into $N$ steps. We index the noise level by the integer $t\in\{1,\dots,N\}$ (continuous time $t/N$), for which the retention coefficient takes the familiar discrete form $\bar\alpha_t=\prod_{i=1}^{t}(1-\beta_i)$ with $\beta_i=\beta(i/N)/N$~\cite{ho2020denoising}.}.

\begin{figure}[t]
    \centering
    \includegraphics[width=\linewidth]{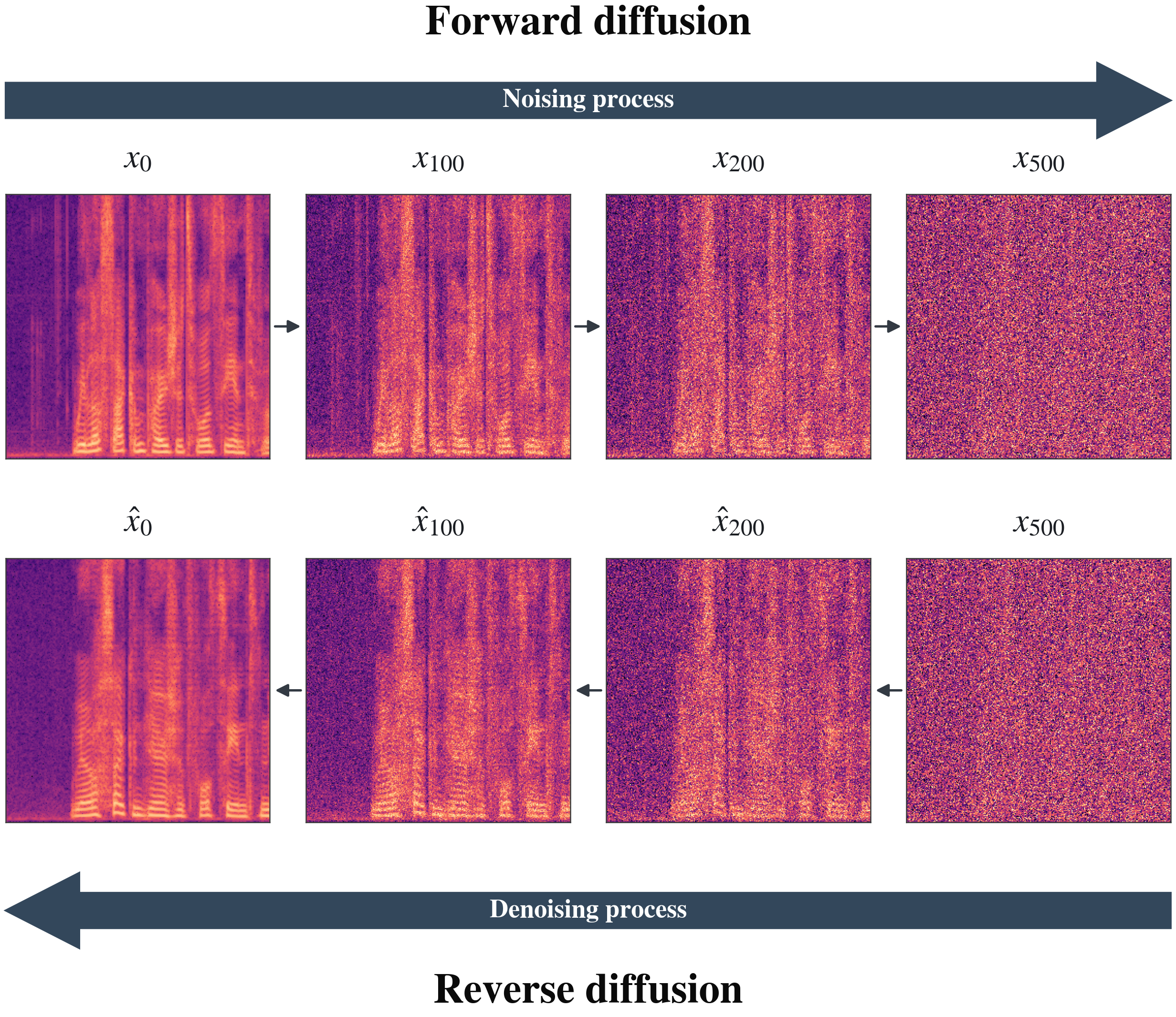}
    \caption{Illustration of the forward and reverse diffusion process.}
    \label{fig:diffusion}
\end{figure}

\noindent\textbf{Reverse process.}
Generation integrates the reverse-time SDE associated with
Eq.~\eqref{eq:sde}~\cite{song2020score,anderson1982reverse} as follows:
\begin{equation}
  \mathrm{d}\hat{x} = \big[f(\hat{x},t) - g(t)^2\,\nabla_{\hat{x}}\log p_t(\hat{x})\big]\mathrm{d}t
  + g(t)\,\mathrm{d}\bar{w},
  \label{eq:revsdegen}
\end{equation}
where $\hat{x}$ is the reverse-time state, $\bar{w}$ is a reverse-time Wiener
process, and $\nabla_{\hat{x}}\log p_t$ is the score of the marginal $p_t$.
Substituting the VP-SDE coefficients $f(\hat{x},t)=-\tfrac{1}{2}\beta(t)\,\hat{x}$ and $g(t)^2=\beta(t)$ into Eq.~\eqref{eq:revsdegen} and replacing the true score $\nabla_{\hat{x}}\log p_t(\hat{x})$ with the learned approximation $s_\theta(\hat{x},t)$ from a trained network gives:
\begin{equation}
  \mathrm{d}\hat{x} =
  \Big[-\tfrac{1}{2}\beta(t)\,\hat{x} - \beta(t)\,s_\theta(\hat{x},t)\Big]\mathrm{d}t
  + \sqrt{\beta(t)}\,\mathrm{d}\bar{w}.
  \label{eq:revsde}
\end{equation}
This SDE is integrated from level $t$ back to $0$, progressively denoising the sample toward the data distribution $p_0$. Fig.~\ref{fig:diffusion} illustrates the forward and reverse trajectories on a representative example.

\subsection{Threat Model}
\label{sec:threat}

\noindent\textbf{Detection setup.}
Let $f_\phi$ denote a fixed, pretrained audio deepfake detector that maps an
input waveform $v$ to a logit vector over the two classes
$\{\textit{spoof},\textit{bonafide}\}$. We summarize its decision by the
bonafide score, the logit assigned to the bonafide class,
\begin{equation}
s(v) = \big[f_\phi(v)\big]_{\textit{bonafide}},
\label{eq:score}
\end{equation}
and classify $v$ as bonafide if $s(v)\ge\tau$ and as spoof otherwise, where
$\tau$ is the threshold chosen at the equal error rate (EER) operating
point. A spoofed utterance $v_s$ is correctly detected when $s(v_s)<\tau$.

\noindent\textbf{Adversary's goal.}
We consider an evasion adversary that aims to bypass the deepfake detector, i.e., to induce a false negative. Specifically, given a spoof $v_s$ that is correctly classified as spoof, $s(v_s)<\tau$, the adversary seeks an additive waveform perturbation $\delta$ such that: 
\begin{equation}
s(v_s+\delta)\ge\tau,
\qquad\text{subject to}\qquad
\lVert\delta\rVert_p\le\epsilon_{\mathrm{adv}}.
\label{eq:goal}
\end{equation}
The constraint bounds the $\ell_p$ norm of the perturbation by the attack budget $\epsilon_{\mathrm{adv}}$, thereby limiting the magnitude of the modification applied to the waveform. We denote the adversarial spoofing audio by $v_a=v_s+\delta$. Rather than synthesizing a new deepfake, the adversary only perturbs an existing spoofed waveform, which keeps the attack inexpensive and the perturbation perceptually small.

\noindent\textbf{Adversary's knowledge and capability.}
We assume a white-box adversary with full access to $f_\phi$, including its
architecture, parameters, and gradients. This worst-case assumption follows Kerckhoffs's principle and ensures the defense is evaluated against the strongest attacks~\cite{carlini2017towards}, and a defense effective in this setting also covers weaker adversaries with partial or no access to the detector. The perturbation is restricted to an $\ell_p$ ball of radius $\epsilon_{\mathrm{adv}}$ with $p\in\{2,\infty\}$ for norm-bounded attacks. We additionally consider norm-minimizing attacks that instead seek the smallest perturbation achieving evasion. In both cases the distortion is kept small. Unless stated otherwise, the adversary is unaware of the defense. The defense-aware adaptive setting is examined separately in Section~\ref{sec:adaptive}.

\noindent\textbf{Defender's setting.}
The defender operates under realistic deployment constraints. The detector
$f_\phi$ is fixed and cannot be retrained or modified. The defender has no
knowledge of the attack type, norm, or budget at test time. Defense is applied
only as an input-level purification step that transforms the waveform before
detection, leaving $f_\phi$ unchanged. This matches the common case of a
deployed detector that must be protected without altering it.

\noindent\textbf{Defender's objective.}
We require the defense to satisfy three competing requirements: (1)~restoring
detection on adversarial spoofing audios; (2)~preserving spoof artifacts, so that a clean (non-adversarial) spoof is not turned into a bonafide-looking sample; and (3)~preserving bonafide speech, leaving it correctly classified.

\section{Methodology}
\label{sec:method}

\subsection{DGAP System Overview}
\label{sec:overview}
\begin{figure*}[t]
    \centering
    \includegraphics[width=\linewidth]{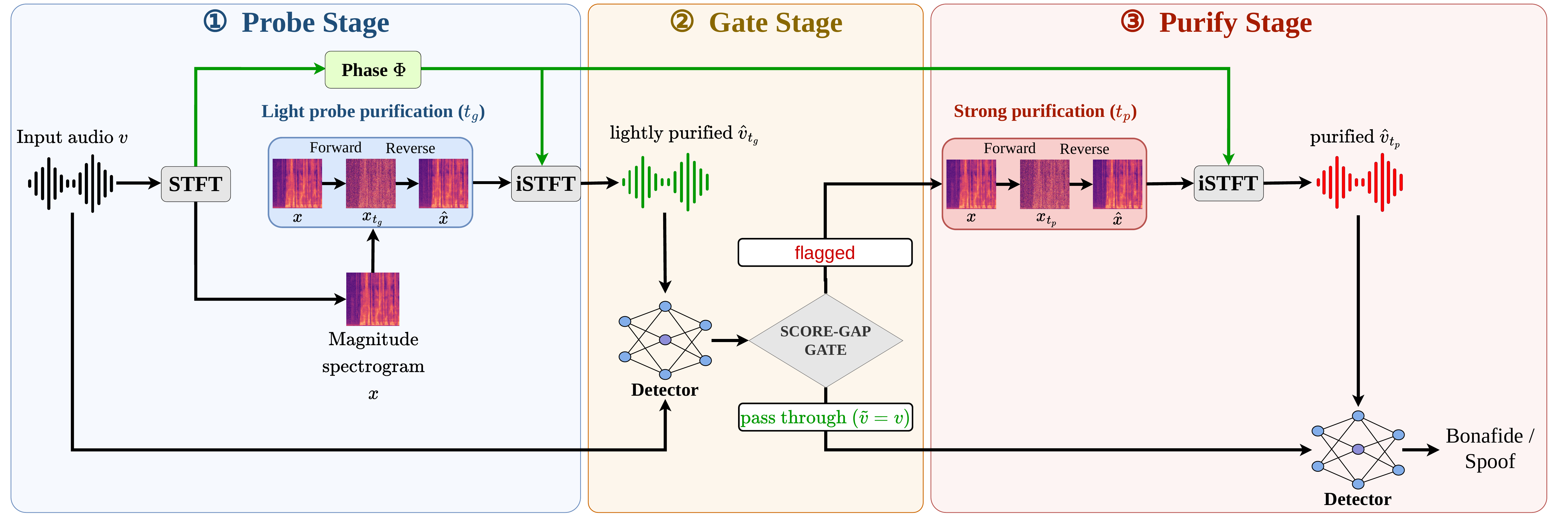}
    \caption{Overview of the proposed DGAP framework.}
    \label{fig:methodology}
\end{figure*}

Our proposed purification framework, 
DGAP, sets the purification strength per input from the detector's own
response, rather than applying a single fixed strength to every utterance, and leaves the pretrained detector unchanged. At its core, a diffusion model fine-tuned on audio spectrograms performs the purification through three stages, namely probe, gate, and purify, as shown in Fig.~\ref{fig:methodology}. In the probe stage, DGAP applies a brief forward-reverse diffusion process at a low noise level to an incoming utterance and measures the resulting shift between the detector scores of the original waveform and the purified copy. Intuitively, a light probe purification perturbs an adversarial input far more than a benign one, so this score gap serves as a per-input, reference-free indicator of adversarial manipulation. Here, a benign input denotes any non-adversarial input, whether bonafide or clean spoof. In the gate stage, an input whose gap exceeds a threshold calibrated on benign inputs is treated as adversarial, whereas an input below the threshold is forwarded unchanged to preserve the discriminative cues the detector exploits. Finally, in the purify stage, only a flagged input is purified at a substantially stronger level, and the resulting waveform is passed to the detector for the final decision.
\subsection{Diffusion-Based Spectrogram Purification for Deepfake Detection}
\label{sec:purifier}

Diffusion-based audio purification methods typically operate either on the raw waveform~\cite{wu2023defending} or on a time-frequency representation such as a spectrogram~\cite{guo2024wavepurifier}. We operate in the spectrogram domain, so that a pretrained image diffusion model can serve as the purification backbone. Specifically, we compute the magnitude spectrogram of each utterance, purify it with the diffusion model, and retain the phase for waveform reconstruction.

Given an input waveform $v\in[-1,1]^L$ of length $L$, we compute its short-time Fourier transform (STFT) using an FFT size $N_{\mathrm{FFT}}$, hop size $H$, and analysis window $\omega$,
\begin{equation}
X(m,k)=\sum_{n=0}^{N_{\mathrm{FFT}}-1}
v[n+mH]\,\omega[n]\,e^{-j2\pi kn/N_{\mathrm{FFT}}},
\label{eq:stft}
\end{equation}
where $m$ indexes time frames and $k$ indexes frequency bins.
We decompose the complex STFT into magnitude and phase,
$A(m,k)=|X(m,k)|$ and $\Phi(m,k)=\angle X(m,k)$.

The diffusion model operates on a decibel-scaled magnitude spectrogram $x$,
linearly normalized to a fixed range, which serves as the diffusion input
$x_0$ in Section~\ref{sec:diffusion}. The decibel scaling compresses
the spectral dynamic range and yields a stable diffusion input, while the phase $\Phi$ is retained for waveform reconstruction. Because the diffusion backbone expects a fixed spatial resolution, the FFT size is chosen so that the frequency axis matches the window size, and $x$ is split along the time axis into square windows that are purified independently and reassembled in their original order. If the utterance is shorter than one window, the spectrogram is padded along the time axis. If the final temporal segment is shorter than the window size, the last window is formed using the available final frames together with neighboring temporal context so that every diffusion input has the required shape. During reassembly, any borrowed temporal context is discarded, and only the frames corresponding to the original utterance are retained. Frequency bins not used by the diffusion model, if any, are retained from the input spectrogram, together with the phase, before inverse STFT.

We fine-tune the diffusion model described in Section~\ref{sec:diffusion} on
spectrogram windows, using the architecture of Dhariwal et al.~\cite{dhariwal2021diffusion}. Unlike speech-recognition purifiers trained only on natural speech~\cite{wu2023defending,guo2024wavepurifier}, we draw the training windows from both bonafide and spoofed utterances, so the learned distribution covers both classes rather than bonafide speech alone.

At inference, the fine-tuned model acts as a window-level purification operator, following the diffusion purification scheme of DiffPure~\cite{nie2022diffusion}. Let $x^{(r)}$ denote the $r$-th spectrogram window. Given a noise level $t$, DGAP diffuses each window to $x_t^{(r)}$ with the forward kernel of Eq.~\eqref{eq:kernel} and integrates the reverse-time SDE of Eq.~\eqref{eq:revsde} back to zero,
\begin{equation}
\begin{aligned}
\hat{x}^{(r)}
&= P_t\big(x^{(r)}\big)
 = \mathrm{sdeint}\big(x_t^{(r)},\, t\!\to\!0\big),\\
x_t^{(r)}
&= \sqrt{\bar\alpha_t}\,x^{(r)}
 + \sqrt{1-\bar\alpha_t}\,\epsilon,
\end{aligned}
\label{eq:purify}
\end{equation}

where $\epsilon\sim\mathcal{N}(0,I)$. The forward step injects Gaussian noise that overwhelms small, off-manifold adversarial perturbations, and the reverse step regenerates each spectrogram window consistently with the learned data distribution. The purified windows $\{\hat{x}^{(r)}\}_r$ are then reassembled in their original temporal order to form the full purified spectrogram $\hat{x}$. 

The purified spectrogram $\hat{x}$ is then inverse-normalized and mapped back to a linear magnitude spectrogram, denoted by $\hat{A}$. The purified magnitude is then recombined with the retained phase:
\begin{equation}
\hat{X}(m,k)=\hat{A}(m,k)\,e^{j\Phi(m,k)}.
\label{eq:recombine}
\end{equation}
The purified waveform at noise level $t$ is obtained by applying the inverse STFT:
\begin{equation}
\hat{v}_t=\operatorname{iSTFT}(\hat{X}),
\label{eq:istft}
\end{equation}
and the unchanged detector is evaluated on $\hat{v}_t$.

The noise level $t$ controls the purification strength through the retention
coefficient $\bar\alpha_t$. A large $t$ suppresses more of the perturbation but also removes input-specific detail and may distort the fine synthesis artifacts the detector relies on. A small $t$ preserves these artifacts more faithfully but may leave part of the perturbation intact. No single fixed $t$ satisfies requirements~(1)--(3) of Section~\ref{sec:threat} across all inputs, which motivates selecting the strength per input from the detector's response, as described in the next section.

\subsection{Detection-Guided Adaptive Gating}
\label{sec:gating}

DGAP decides how much to purify each input from the detector's own response. The detector operates on waveforms, whereas the learned purifier $P_t$ operates on magnitude spectrogram windows. For an input waveform $v$, we write $\hat{v}_t(v)$ for its purified reconstruction at noise level $t$, as defined in Section~\ref{sec:purifier}.

Using the bonafide score $s$ of Eq.~\eqref{eq:score}, DGAP probes an input
waveform $v$ at a small level $t_g$ and measures the induced shift in the detector score:
\begin{equation}
d_{t_g}(v) = s(v) - s\big(\hat{v}_{t_g}(v)\big),
\label{eq:gate}
\end{equation}
which we call the score gap. It is computed from the detector's outputs alone and needs no clean reference for the input under test, so it is available at inference.

The gap separates adversarial from benign inputs because a light purification
changes the detector response differently in the two cases. Let $v_s$ be a clean spoof and $v_a=v_s+\delta$ the corresponding adversarial spoof,
with $\lVert\delta\rVert_p\le\epsilon_{\mathrm{adv}}$ as defined in
Section~\ref{sec:threat}. Let $T(\cdot)$ denote the mapping from a waveform to its normalized decibel-magnitude spectrogram representation. We write
$x_s=T(v_s)$ and $x_a=T(v_a)$, and define the representation-domain difference $\Delta_x=x_a-x_s$. If both representations are diffused to level $t_g$ under a common noise draw, applied window-wise as in Eq.~\eqref{eq:purify}, their noised states $x_{a,t_g}$ and $x_{s,t_g}$ satisfy
\begin{equation}
\big\lVert x_{a,t_g}-x_{s,t_g}\big\rVert
=
\sqrt{\bar\alpha_{t_g}}\,\lVert\Delta_x\rVert,
\label{eq:contract}
\end{equation}
where $\lVert\cdot\rVert$ denotes a norm in the diffusion representation space. Since $\bar\alpha_t$ is non-increasing in $t$, Eq.~\eqref{eq:contract} implies that, for $t_2>t_1$,
\begin{equation}
\big\lVert x_{a,t_2}-x_{s,t_2}\big\rVert
\le
\big\lVert x_{a,t_1}-x_{s,t_1}\big\rVert .
\label{eq:monotone_contract}
\end{equation}
Thus, in the forward noising stage, larger diffusion levels attenuate
representation-domain adversarial differences more strongly. After reverse
denoising and waveform reconstruction, this attenuation can appear as a change
in the detector's bonafide score. The score gap $d_{t_g}$ measures this change
at the detector output.

For a successful attack, the adversarial spoof $v_a$ is misclassified as bonafide before purification, so $s(v_a)\ge\tau$. After the light probe purification, its detector score is expected to move back toward the spoof side, so that
\begin{equation}
d_{t_g}(v_a) = s(v_a)-s\big(\hat{v}_{t_g}(v_a)\big) > 0.
\label{eq:gapdir}
\end{equation}
In contrast, a benign input is already close to the learned data distribution,
so a light purification changes it only slightly, giving $\hat{v}_{t_g}(v)\approx v$ and $d_{t_g}(v)\approx 0$.

We turn this signal into a decision by flagging an input whose gap is anomalously large relative to benign inputs. On a calibration set of benign utterances, we record $d_{t_g}$ and set the threshold: 
\begin{equation}
\eta = \mu_c + \gamma\,\sigma_c,
\label{eq:eta}
\end{equation}
where $\mu_c$ and $\sigma_c$ are the mean and standard deviation of $d_{t_g}$ on the calibration set, and $\gamma$ fixes the operating point. At inference, the input is strongly purified only when flagged,
\begin{equation}
\tilde{v} =
\begin{cases}
\hat{v}_{t_p}(v), & d_{t_g}(v) > \eta \quad(\text{flagged}),\\[2pt]
v,              & \text{otherwise}.
\end{cases}
\label{eq:policy}
\end{equation}
The unchanged detector is then applied to $\tilde{v}$. Algorithm~\ref{alg:dgap} illustrates the complete inference procedure. The probe level $t_g$ and the purification level $t_p$ are decoupled, with $t_g$ exposing the gap and $t_p$ chosen to remove the perturbation when purification is required.

\begin{algorithm}[t]
\caption{DGAP inference}
\label{alg:dgap}
\begin{algorithmic}[1]
\Require input waveform $v$; detector score $s$, detector threshold $\tau$;
spectrogram-window purifier $P_t$; probe/purify levels $t_g,t_p$;
gate threshold $\eta$
\State Construct $\hat{v}_{t_g}(v)$ by applying $P_{t_g}$ to the spectrogram windows of $v$ and reconstructing the waveform
\State $d_{t_g} \gets s(v) - s\big(\hat{v}_{t_g}(v)\big)$
\Comment{probe: score gap, Eq.~\eqref{eq:gate}}
\If{$d_{t_g} > \eta$}
  \Comment{gate, Eq.~\eqref{eq:policy}}
  \State Construct $\hat{v}_{t_p}(v)$ by applying $P_{t_p}$ to the spectrogram windows of $v$ and reconstructing the waveform
  \State $\tilde{v} \gets \hat{v}_{t_p}(v)$
  \Comment{strong purification}
\Else
  \State $\tilde{v} \gets v$
  \Comment{pass through}
\EndIf
\State \Return $\textsc{bonafide}$ if $s(\tilde{v}) \ge \tau$ else $\textsc{spoof}$
\end{algorithmic}
\end{algorithm}

The parameters $(t_g,t_p,\gamma)$ are selected on a held-out development split by minimizing the objective:
\begin{equation}
y=
\mathrm{EER}\big(s(\tilde{v}),\,\mathcal{D}_{\mathrm{clean}}\big)
+\lambda\,\mathrm{EER}\big(s(\tilde{v}),\,\mathcal{D}_{\mathrm{adv}}\big),
\label{eq:objective}
\end{equation}
where $\tilde{v}$ is the DGAP output in Eq.~\eqref{eq:policy},
$s(\tilde{v})$ is the detector score from Eq.~\eqref{eq:score},
$\mathcal{D}_{\mathrm{clean}}$ contains non-adversarial bonafide and spoofed
utterances, and $\mathcal{D}_{\mathrm{adv}}$ contains bonafide and adversarial
spoofed utterances. Here, $\mathrm{EER}\big(s(\tilde{v}),\mathcal{D}\big)$
denotes the equal error rate of the scores $s(\tilde{v})$ over $\mathcal{D}$. The first term penalizes degradation on benign utterances, while the second penalizes residual vulnerability on adversarial spoofs. The coefficient $\lambda$ balances clean-data preservation and attack removal.

\section{Evaluation}
\subsection{Experiment Settings}
\label{sec:setup}

\noindent\textbf{Target detection models.}
To evaluate the proposed defense framework, we use three publicly available
pretrained detectors for the ASVspoof~2019 LA task, namely
AASIST~\cite{jung2022aasist}, RawGAT-ST~\cite{tak2021end1}, and
Res-TSSDNet~\cite{hua2021towards}. The first two
are graph-attention detectors that operate on raw waveforms and model
spectro-temporal dependencies through attention. In contrast, Res-TSSDNet adopts a fully time-domain residual architecture and classifies directly from waveform samples without an explicit spectral front-end. Each detector maps an input waveform to a bonafide score $s$. We use the pretrained checkpoints for all three detectors and keep them frozen throughout evaluation.

\noindent\textbf{Audio adversarial attacks.}
We evaluate our method against three white-box adversarial attack
settings, all realizing the evasion goal of Eq.~\eqref{eq:goal}. PGD~\cite{madry2017towards}, a norm-bounded attack, iteratively minimizes the cross-entropy toward the bonafide target within an $\ell_p$ ball. We use its $\ell_\infty$ ($\epsilon_{\mathrm{adv}}=0.005$) and $\ell_2$ ($\epsilon_{\mathrm{adv}}=1.0$) variants, with the budget held fixed across detectors and the number of iterations tuned per detector to reach a comparable success rate. The Carlini--Wagner (C\&W)~\cite{carlini2017towards} attack, a norm-minimizing attack, instead minimizes the perturbation norm under a confidence-margin penalty, with confidence $\kappa=10$, balancing constant $c=1.0$, and $200$ optimization steps. Table~\ref{tab:asr} reports the resulting attack success rates on the undefended detectors.
\begin{table}[t]
\centering
\caption{White-box attack success rate (\%) on the undefended detectors, with the
number of PGD iterations in parentheses.}
\label{tab:asr}
\begin{tabular}{lccc}
\toprule
Detector & PGD-$\ell_\infty$ & PGD-$\ell_2$ & C\&W \\
\midrule
AASIST       & 94.0 \,(20)  & 94.5 \,(10) & 95.0 \\
RawGAT-ST    & 73.2 \,(100) & 91.2 \,(50) & 97.9 \\
Res-TSSDNet  & 82.5 \,(20)  & 97.8 \,(10) & 100.0 \\
\bottomrule
\end{tabular}
\end{table}

\noindent\textbf{Baselines.}
We compare our method with widely used defense methodologies for audio adversarial robustness, adapted to the audio deepfake detection setting. The baselines cover three defense families. Transformation-based defenses apply fixed waveform preprocessing. Following WaveGuard~\cite{hussain2021waveguard}, we include quantization (QT), downsampling-upsampling (DS), and linear predictive coding (LPC), together with low-pass and band-pass filtering (LPF, BPF)~\cite{rajaratnam2018noise}. Noise-based defenses add noise to the input before detection, for which we use inference-time Gaussian noise addition (ITNA)~\cite{kim2025inference} and randomized smoothing (RS)~\cite{cohen2019certified}. Finally, as diffusion-based purification baselines, we use DiffPure~\cite{nie2022diffusion} and WavePurifier~\cite{guo2024wavepurifier}, both implemented with the same fine-tuned spectrogram purifier as our method and purifying every input at a fixed, input-agnostic strength. DiffPure applies a single diffusion level and serves as the uniform baseline, whereas WavePurifier follows a hierarchical band-split schedule.

\noindent\textbf{Dataset and adversarial samples.}
Our method is evaluated on the logical access (LA) partition of the ASVspoof~2019 dataset~\cite{todisco2019asvspoof}, a widely used benchmark for audio deepfake detection. The partition contains $16$\,kHz bonafide and spoofed utterances of varying duration, with spoofed samples generated by text-to-speech and voice-conversion systems. The dataset provides official training, development, and evaluation splits with $25{,}380$, $24{,}844$, and $71{,}237$ utterances, respectively. To generate adversarial examples, we use $600$ correctly classified spoofed utterances from the evaluation split for testing and $500$ from the development split for parameter selection, each paired with an equal number of bonafide utterances. On each split, these utterances instantiate $\mathcal{D}_{\mathrm{clean}}$, and replacing the clean spoofs with their adversarial versions yields $\mathcal{D}_{\mathrm{adv}}$ of Eq.~\eqref{eq:objective}. The underlying utterance sets are kept identical across detectors. The calibration statistics $(\mu_c,\sigma_c)$ of Eq.~\eqref{eq:eta} are computed on the development $\mathcal{D}_{\mathrm{clean}}$.

\noindent\textbf{Diffusion model.}
We use OpenAI's guided-diffusion model~\cite{dhariwal2021diffusion}, initialized from its publicly available $256\times256$ pretrained checkpoint, with the noise schedule discretized into $N=1000$ steps. We fine-tune it on $55{,}001$ spectrogram windows of the corresponding size extracted from training utterances. Spectrograms are computed at a $16$\,kHz sampling rate using a $512$-point STFT, a hop size of $128$, and a Hann window. The magnitude is mapped to a decibel scale and shifted by a $-20$\,dB offset, then min-max normalized from $[-120,25]$\,dB to $[0,1]$.

\noindent\textbf{Metrics.}
We report the EER, the standard metric in anti-spoofing. The EER is the value at which the false acceptance and false rejection rates are equal, obtained by sweeping the decision threshold. Under a defense, both bonafide and spoofed utterances are passed through the defense before scoring, and the EER is computed from the resulting detector scores. $\mathrm{EER}_{\mathrm{clean}}$ (abbreviated $\mathrm{EER}_{c}$ in tables) is measured on $\mathcal{D}_{\mathrm{clean}}$ and $\mathrm{EER}_{\mathrm{adv}}$ on $\mathcal{D}_{\mathrm{adv}}$, quantifying clean condition discrimination and residual vulnerability after defense, respectively. We report $\mathrm{EER}_{\mathrm{adv}}$ separately for each attack, and the attack columns in the tables (PGD-$\ell_\infty$, PGD-$\ell_2$, and C\&W) denote the $\mathrm{EER}_{\mathrm{adv}}$ measured under the corresponding attack. A strong defense achieves a low $\mathrm{EER}_{\mathrm{adv}}$ while keeping $\mathrm{EER}_{\mathrm{clean}}$ close to its no-defense value, a trade-off captured by the objective $y$ of Eq.~\eqref{eq:objective}.

\begin{figure*}[t]
\centering
\includegraphics[width=0.75\textwidth]{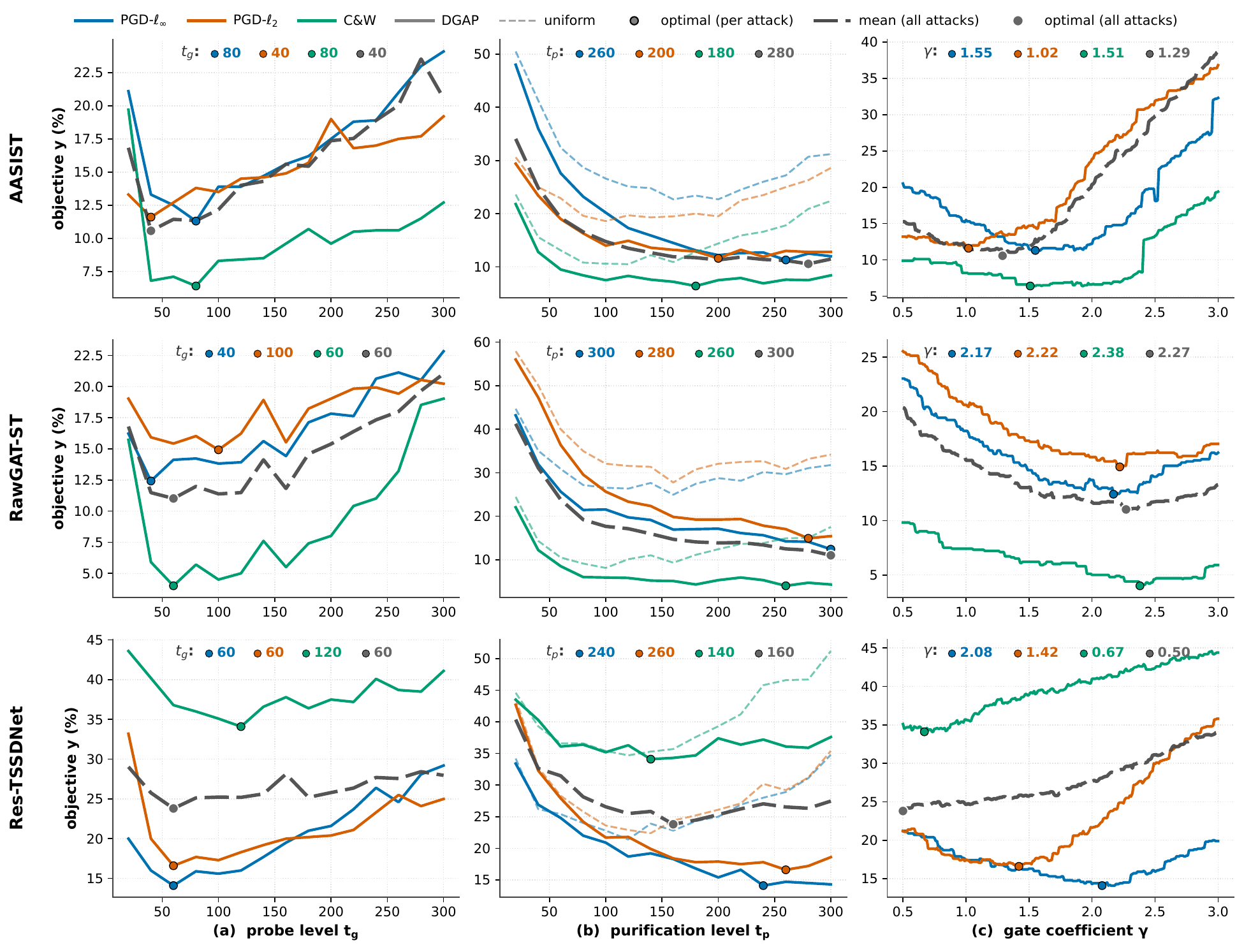}
\caption{Parameter selection on the development split. Each panel plots the objective $y$ versus (a) the probe level $t_g$, (b) the purification level $t_p$, and (c) the gate coefficient $\gamma$. Panel (b) also shows the uniform baseline. Filled markers denote per-attack optima, the grey dashed curve is the objective averaged over the three attacks, and the grey marker denotes the single attack-agnostic configuration selected per detector.}
\label{fig:param_selection}
\end{figure*}

\subsection{Selecting the Purification Parameters}
\label{sec:select}
We select the purification parameters on the development split separately for each combination of detector and attack. We jointly search a grid of $15$ diffusion levels from $20$ to $300$ in steps of $20$ for both the probe level $t_g$ and the purification level $t_p$, together with gate coefficients $\gamma\in[0.5,3.0]$ at intervals of $0.01$. From this grid, we identify the configuration $(t_g,t_p,\gamma)$ that minimizes the development objective of Eq.~\eqref{eq:objective}. We set $\lambda = 1$, giving equal weight to clean-data preservation and adversarial robustness. Fig.~\ref{fig:param_selection} visualizes this search, varying one parameter at a time with the other two fixed at their optimum.

We first examine the effect of the probe level $t_g$ on the development objective, as shown in Fig.~\ref{fig:param_selection}(a). Across detectors and attacks, we observe that the objective is minimized at relatively small probe levels. This suggests that a light probe is sufficient to reveal the score gap induced by adversarial perturbations. Increasing $t_g$ beyond this range degrades the gap signal because stronger probing also modifies benign inputs and reduces the contrast between benign and adversarial utterances.

We next examine how the purification level $t_p$ behaves, as shown in Fig.~\ref{fig:param_selection}(b). For the uniform baseline, the optimal level reflects the expected compromise between attack removal and clean-data preservation. Moderate purification reduces adversarial vulnerability, whereas larger diffusion levels begin to degrade clean performance by attenuating synthesis artifacts that are useful for detection. In contrast, our method's optimum lies at a stronger purification level, because it applies this level only to inputs flagged by the gate. As a result, its objective continues to improve or remains stable at larger $t_p$ values, while the uniform baseline reaches its optimum at a lower level.

Finally, we analyze the effect of the gate coefficient $\gamma$, as reported in Fig.~\ref{fig:param_selection}(c). We find that small values of $\gamma$ produce an aggressive gate that may unnecessarily purify benign inputs, whereas large values produce a conservative gate that may leave some adversarial inputs unpurified. The optimal value therefore balances clean-data preservation and attack removal. The optimal $\gamma$ values follow the separability of the score-gap distributions. Pairs with stronger separation can use a more conservative gate, whereas harder cases require a lower threshold.

\subsection{Main Results}

\begin{figure}[t]
    \centering
    \includegraphics[width=\linewidth]{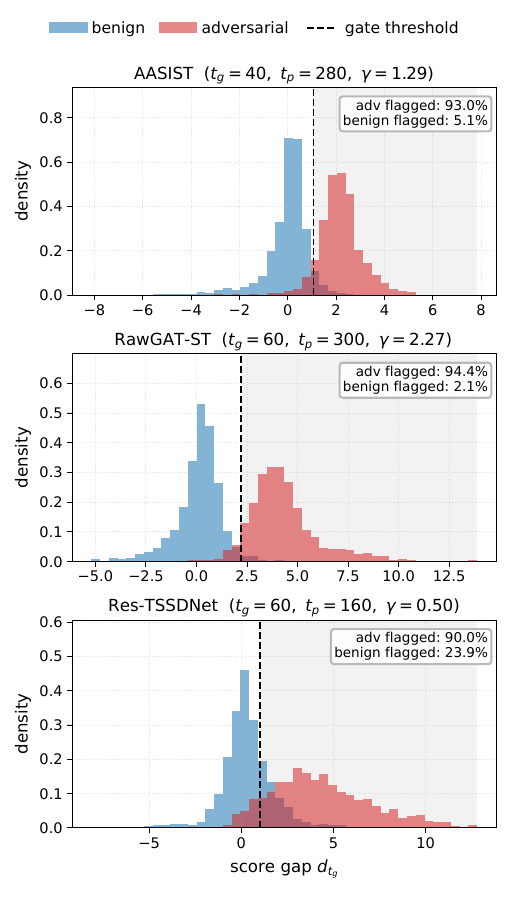}
    \caption{Score-gap distributions at the probe level $t_g$ for benign and adversarial inputs.}
    \label{fig:gate_threshold}
\end{figure}

As the per-attack optima in Fig.~\ref{fig:param_selection} indicate, the configuration that is optimal for one attack is generally suboptimal for another. To obtain a defense that does not rely on prior knowledge of the attack, we use a single attack-agnostic configuration for each detector across all attacks. Specifically, we select the configuration that achieves the lowest development objective averaged over the three attacks, as shown in Fig.~\ref{fig:param_selection}. Retuning per attack would instead require knowledge of the attack type, norm, and perturbation budget at test time, which is unavailable to the defender under our threat model. Unless otherwise specified, all subsequent experiments use this attack-agnostic configuration. Fig.~\ref{fig:gate_threshold} shows how the selected setting separates benign from adversarial inputs on the evaluation split, pooled over the three attacks. At the selected probe level $t_g$, purification perturbs the scores of benign inputs only marginally but shifts those of adversarial inputs substantially, yielding score-gap distributions with limited overlap. The gate threshold consequently flags over $90\%$ of adversarial utterances for the strong purification level $t_p$ while leaving the large majority of benign inputs unmodified.

Table~\ref{tab:overall} reports the clean and adversarial EER and the combined objective $y$ on the evaluation split. The undefended detectors are highly vulnerable, with $\mathrm{EER}_{\mathrm{adv}}$ reaching $61.92\%$ on AASIST and $57.49\%$ on Res-TSSDNet. The transformation-based defenses recover only partially and inconsistently across attacks. The strongest of them still leaves $y$ at $9.58\%$ on AASIST and $16.76\%$ on RawGAT-ST. On several attacks a
transform is even worse than no defense at all. For instance, DS, LPF, and the two noise-based defenses all raise the RawGAT-ST C\&W EER above its undefended value of $20.70\%$. This happens because these defenses apply a fixed, input-agnostic distortion, and when the residual attack signal is already small they corrupt the discriminative cues more than they remove the perturbation. The noise-based defenses further pay a heavy clean-data cost, inflating $\mathrm{EER}_{\mathrm{clean}}$ to between $9.83\%$ and $14.42\%$ on AASIST and above $21\%$ on Res-TSSDNet, since the injected noise also distorts benign inputs. The two diffusion baselines are the most consistent across detectors, yet neither resolves the core trade-off. Because they purify every input at the same strength, they cannot reconcile robustness with clean preservation. The hierarchical,
band-split schedule of WavePurifier does not change this picture. It is marginally worse than the uniform baseline (DiffPure) on AASIST and RawGAT-ST and only slightly better on Res-TSSDNet.

Our method attains the best combined objective $y$ on all three detectors, lowering it from the uniform baseline (DiffPure)'s $12.56\%$, $11.26\%$, and $29.69\%$ to $5.61\%$, $5.18\%$, and $21.65\%$, respectively. Because the strong purification level is applied only to gated inputs, benign utterances pass through almost untouched and $\mathrm{EER}_{\mathrm{clean}}$ stays within less than one point of the no-defense value on every detector. The advantage is clearest on AASIST and RawGAT-ST, where our method also achieves the lowest
$\mathrm{EER}_{\mathrm{adv}}$ in nearly every column. On Res-TSSDNet, the hardest detector to defend, the margin narrows and individual single-attack columns are occasionally led by a transformation-based defense. These methods, however, fail to generalize across conditions. For instance, QT is competitive under $\ell_\infty$ and $\ell_2$ but collapses to $54.29\%$ under C\&W and degrades $\mathrm{EER}_{\mathrm{clean}}$ to $8.33\%$, whereas our defense remains balanced across all attacks and yields the best $y$ on this detector as well.

\begin{table*}[t]
\centering
\caption{$\mathrm{EER}_{c}$ and per-attack $\mathrm{EER}_{\mathrm{adv}}$ (\%) of the defense methods, and the objective $y$ with the adversarial term averaged over the three attacks. Best in \textbf{bold}, second best \underline{underlined}.}
\label{tab:overall}
\scriptsize
\setlength{\tabcolsep}{3.2pt}
\renewcommand{\arraystretch}{1.08}
\resizebox{\textwidth}{!}{
\begin{tabular}{l|ccccc|ccccc|ccccc}
\toprule
& \multicolumn{5}{c|}{AASIST}
& \multicolumn{5}{c|}{RawGAT-ST}
& \multicolumn{5}{c}{Res-TSSDNet} \\
\cmidrule(lr){2-6}
\cmidrule(lr){7-11}
\cmidrule(lr){12-16}
Method
& $\mathrm{EER}_{c}$ & $\ell_{\infty}$ & $\ell_{2}$ & C\&W & $y$
& $\mathrm{EER}_{c}$ & $\ell_{\infty}$ & $\ell_{2}$ & C\&W & $y$
& $\mathrm{EER}_{c}$ & $\ell_{\infty}$ & $\ell_{2}$ & C\&W & $y$ \\
\midrule
No defense
& 1.17 & 61.92 & 49.42 & 46.50 & 53.78
& 1.25 & 21.04 & 26.47 & 20.70 & 23.99
& 2.17 & 57.49 & 50.08 & 38.47 & 50.85 \\
\midrule
QT
& 4.08 & \textbf{4.08} & \underline{5.42} & 7.00 & \underline{9.58}
& 2.17 & 9.16 & 19.26 & 15.35 & 16.76
& 8.33 & \textbf{17.43} & \textbf{12.71} & 54.29 & 36.48 \\
DS
& 2.67 & 12.75 & 13.08 & 22.25 & 18.70
& 2.67 & 20.36 & 26.47 & 33.51 & 29.45
& \underline{3.00} & 29.71 & 34.18 & 30.64 & 34.51 \\
LPF
& \underline{2.08} & 12.17 & 14.83 & 32.83 & 22.02
& \underline{1.83} & 17.56 & 22.31 & 23.66 & 23.01
& 4.50 & 26.94 & 34.18 & 36.36 & 36.99 \\
BPF
& 2.58 & 8.75 & 9.58 & 19.50 & 15.19
& \underline{1.83} & 11.71 & 17.56 & 16.20 & 16.99
& 3.08 & 30.22 & 32.24 & 28.11 & 33.27 \\
LPC
& 2.75 & 48.92 & 34.33 & 29.25 & 40.25
& \underline{1.83} & 20.18 & 26.79 & 15.78 & 22.75
& 4.50 & 46.46 & 44.36 & 47.22 & 50.51 \\
\midrule
ITNA
& 9.83 & 18.83 & 20.58 & 41.42 & 36.77
& \textbf{1.67} & 9.92 & 20.10 & 37.92 & 24.31
& 24.42 & 28.62 & 28.62 & 32.66 & 54.39 \\
RS
& 14.42 & 18.42 & 19.17 & 25.00 & 35.28
& \underline{1.83} & 10.35 & 19.68 & 36.98 & 24.17
& 21.58 & 25.42 & 26.09 & 30.56 & 48.94 \\
\midrule
DiffPure
& 4.11 & 9.17 & 10.42 & 5.75 & 12.56
& 3.17 & 9.25 & \underline{12.04} & \underline{2.97} & \underline{11.26}
& 4.50 & 26.26 & 21.46 & 27.86 & 29.69 \\
WavePurifier
& 3.83 & 12.08 & 10.00 & \underline{5.67} & 13.08
& 3.50 & \underline{9.07} & 12.13 & 3.48 & 11.73
& 4.33 & 23.32 & 20.20 & \underline{27.44} & \underline{27.99} \\
\midrule
\textbf{Ours}
& \textbf{1.83} & \underline{4.25} & \textbf{5.00} & \textbf{2.08} & \textbf{5.61}
& \textbf{1.67} & \textbf{4.33} & \textbf{4.58} & \textbf{1.61} & \textbf{5.18}
& \textbf{2.83} & \underline{19.37} & \underline{15.68} & \textbf{21.40} & \textbf{21.65} \\
\bottomrule
\end{tabular}
}
\end{table*}

\subsection{Ablation Study}
\label{sec:ablation}
To isolate the role of the gate, we compare three configurations on the development split: uniform purification at its per-detector optimal level, corresponding to the DiffPure baseline in Table~\ref{tab:overall}, uniform purification at the stronger level $t_p$ used by the gated method, and gated purification at the same $t_p$, as reported in Table~\ref{tab:ablation}. The comparison between the last two isolates the gate, since they differ only in whether $t_p$ is applied to every input or only to flagged ones.

The results show that the gate enables a purification level that is too aggressive when applied uniformly. Increasing the uniform level from its optimum to $t_p$ worsens $\mathrm{EER}_{\mathrm{adv}}$ in all nine cases and also increases $\mathrm{EER}_{\mathrm{clean}}$, since the indiscriminate purification alters benign utterances and removes discriminative spoofing cues. By reserving $t_p$ for inputs with an anomalously large probe-induced score shift, the gate recovers this degradation and achieves the lowest $\mathrm{EER}_{\mathrm{adv}}$ in every case. The main contribution of the gate is thus decoupling strong attack removal from the distortion imposed on benign inputs. Consistent with this interpretation, the gain is smallest on Res-TSSDNet, where the selected $t_p$ is already close to the optimal uniform level.
\begin{table}[t]
\centering
\caption{Ablation of the gate on the development split: $\mathrm{EER}_{c}$ and per-attack $\mathrm{EER}_{\mathrm{adv}}$ (\%). Uniform purification at the per-detector optimal level (equivalent to DiffPure), uniform purification at the stronger level $t_p$, and the gated method at the same $t_p$. Best in \textbf{bold}.}
\label{tab:ablation}
\scriptsize
\setlength{\tabcolsep}{6pt}
\renewcommand{\arraystretch}{1.08}
\begin{tabular}{llccccc}
\toprule
Detector & Configuration & $t$ & $\mathrm{EER}_{c}$ & $\ell_{\infty}$ & $\ell_{2}$ & C\&W \\
\midrule
\multirow{3}{*}{AASIST}
& Uniform, optimal $t$   & 160 & \textbf{1.90} & 20.80 & 17.60 & 9.00 \\
& Uniform at $t_p$       & 280 & 6.30 & 24.40 & 20.00 & 14.60 \\
& Gated at $t_p$ (Ours)  & 280 & 2.30 & \textbf{9.30} & \textbf{10.30} & \textbf{5.20} \\
\midrule
\multirow{3}{*}{RawGAT-ST}
& Uniform, optimal $t$   & 160 & \textbf{1.30} & 23.65 & 26.45 & 8.02 \\
& Uniform at $t_p$       & 300 & 4.00 & 27.76 & 30.16 & 13.53 \\
& Gated at $t_p$ (Ours)  & 300 & \textbf{1.30} & \textbf{11.72} & \textbf{14.53} & \textbf{2.91} \\
\midrule
\multirow{3}{*}{Res-TSSDNet}
& Uniform, optimal $t$   & 120 & 1.50 & 19.90 & 21.40 & 33.20 \\
& Uniform at $t_p$       & 160 & 1.50 & 21.30 & 22.90 & 34.20 \\
& Gated at $t_p$ (Ours)  & 160 & \textbf{1.10} & \textbf{17.70} & \textbf{17.00} & \textbf{29.20} \\
\bottomrule
\end{tabular}
\end{table}

\subsection{Adaptive Attack Evaluation}
\label{sec:adaptive}

We further evaluate the adaptive attack scenario. Since backpropagating through the
reverse-time SDE is computationally demanding, we follow the BPDA+EOT
protocol used to evaluate stochastic purification
defenses~\cite{athalye2018obfuscated,nie2022diffusion}. Backward Pass
Differentiable Approximation (BPDA) treats the defense as the identity
in the backward pass, and Expectation over Transformation
(EOT)~\cite{athalye2018synthesizing} averages gradients over stochastic evaluations of the full probe, gate, and purify pipeline, so the gate decision is inside the attack loop. The adversary solves Eq.~\eqref{eq:goal} with the defended score in
place of the undefended one, maximizing the expected bonafide score of
the DGAP output $\tilde{v}$ over the randomness of the defense. The attack uses $\ell_\infty$ with $\epsilon_{\mathrm{adv}}{=}0.005$
as in the non-adaptive setting. Because every EOT sample requires a full execution of the purification pipeline, the cost of the adaptive attack grows rapidly with the number of PGD iterations and EOT samples. To balance computational cost against attack strength, we use $20$ PGD iterations and $8$ EOT samples per iteration, evaluate on $120$ spoofed and $120$ bonafide utterances, and report results on AASIST and RawGAT-ST. Each baseline is attacked through its own forward pass, using the same identity approximation for the transformation defenses~\cite{hussain2021waveguard}, for which the attack reduces to BPDA alone since they are deterministic. RS is instead evaluated as a smoothed classifier, with the detector score averaged over $100$ noisy copies of the input. We omit the WavePurifier for adaptive evaluation due to its cost of three diffusion passes per defense evaluation. It is expected to behave like the DiffPure baseline for the deepfake detection task, which it marginally underperforms in the non-adaptive setting (Table~\ref{tab:overall}).

Table~\ref{tab:adaptive} reports the results. On RawGAT-ST, our method
remains the strongest defense, with the best $y$ of $14.17\%$ against
DiffPure's $18.33\%$ and the best transformation's $15.83\%$. On
AASIST, QT takes the lead with $y{=}20.00\%$, while our method ties
with BPF for second at $25.00\%$, ahead of DiffPure ($29.17\%$). The
advantage over DiffPure comes from both terms of the objective. Since
benign inputs are mostly passed through rather than purified,
$\mathrm{EER}_{\mathrm{clean}}$ stays at $0.00\%$ and $0.83\%$, well
below DiffPure's cost of purifying every input. On the adversarial
side, evading the gate constrains the perturbation in addition to
fooling the detector, and the flagged inputs are purified at a
stronger level than the uniform baseline affords. The undefended
$\mathrm{EER}_{\mathrm{adv}}$ is considerably lower on RawGAT-ST than
on AASIST, consistent with the larger number of PGD iterations this
detector required in the non-adaptive setting (Table~\ref{tab:asr}).
On RawGAT-ST, DS and LPC provide no net benefit, matching the $y$ of the undefended detector exactly.

\begin{figure}[t]
    \centering
    \includegraphics[width=\linewidth]{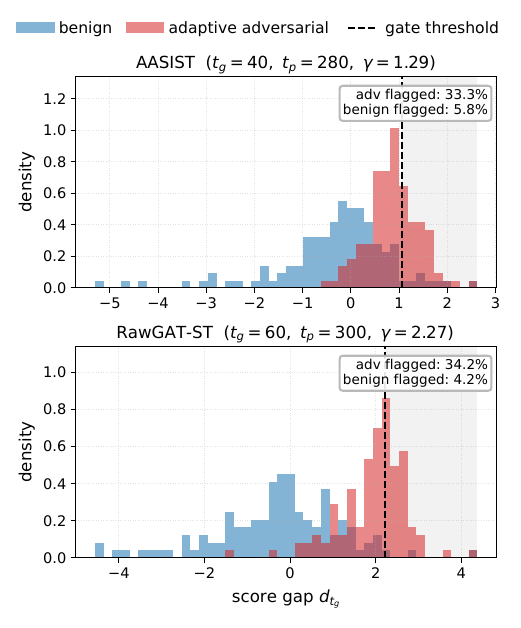}
    \caption{Score-gap distributions under the adaptive BPDA+EOT attack.}
    \label{fig:gate_adaptive}
\end{figure}

The gate statistics explain the gap to the non-adaptive setting. Under the defense-aware attack, only $33.3\%$ and $34.2\%$ of adversarial utterances are flagged, against $5.8\%$ and $4.2\%$ of benign inputs. As shown in Fig.~\ref{fig:gate_adaptive}, the adversary concentrates the score gap of adversarial inputs just below the threshold $\eta$, so that the probe barely shifts the detector score and the strong purification is bypassed. This constraint is not free, as the perturbation must evade the detector while remaining invisible to the probe, which caps the attack well below the undefended level. Gate evasion is thus the principal failure mode of our method under adaptive attack. We note that the reported robustness is conditional on the attack budget, and
adaptive attacks with more PGD steps or larger EOT samples may widen this gap further.
\begin{table}[t]
\centering
\caption{$\mathrm{EER}_{c}$, $\mathrm{EER}_{\mathrm{adv}}$, and the objective $y$ (\%) under the adaptive BPDA+EOT attack. Best in \textbf{bold}, second best \underline{underlined}.}
\label{tab:adaptive}
\scriptsize
\setlength{\tabcolsep}{4.5pt}
\renewcommand{\arraystretch}{1.05}
\begin{tabular}{l|ccc|ccc}
\toprule
& \multicolumn{3}{c|}{AASIST} & \multicolumn{3}{c}{RawGAT-ST} \\
\cmidrule(lr){2-4}\cmidrule(lr){5-7}
Method & $\mathrm{EER}_{c}$ & $\mathrm{EER}_{\mathrm{adv}}$ & $y$
       & $\mathrm{EER}_{c}$ & $\mathrm{EER}_{\mathrm{adv}}$ & $y$ \\
\midrule
No defense & 0.00 & 64.17 & 64.17 & 0.83 & 23.33 & 24.17 \\
\midrule
QT   & 3.33 & \textbf{16.67} & \textbf{20.00} & \textbf{0.83} & 20.00 & 20.83 \\
DS   & \underline{0.83} & 25.00 & 25.83 & 2.50 & 21.67 & 24.17 \\
LPF  & 1.67 & 30.00 & 31.67 & \underline{1.67} & 19.17 & 20.83 \\
BPF  & 1.67 & \underline{23.33} & \underline{25.00} & \underline{1.67} & \underline{14.17} & \underline{15.83} \\
LPC  & \underline{0.83} & 56.67 & 57.50 & \textbf{0.83} & 23.33 & 24.17 \\
\midrule
ITNA & 8.33 & 26.67 & 35.00 & \underline{1.67} & 20.00 & 21.67 \\
RS   & 9.17 & 32.50 & 41.67 & \underline{1.67} & 20.00 & 21.67 \\
\midrule
DiffPure & 2.50 & 26.67 & 29.17 & 4.17 & \underline{14.17} & 18.33 \\
\midrule
\textbf{Ours} & \textbf{0.00} & 25.00 & \underline{25.00} & \textbf{0.83} & \textbf{13.33} & \textbf{14.17} \\
\bottomrule
\end{tabular}
\end{table}

\subsection{Time Budget Analysis}
\label{sec:timebudget}
To assess the practical cost of the gating mechanism, we also examine the per-utterance running time of the proposed method. Table~\ref{tab:timebudget} reports the running time of each stage together with the flag rates under the non-adaptive attacks, measured on a single NVIDIA RTX A6000 GPU with the attack-agnostic configuration. Every input incurs the light probe purification and scoring, whereas the strong purification is incurred only when the gate flags the input. The expected per-utterance time is therefore the probe cost plus the strong-purification cost weighted by the flag rate, reported as the benign and adversarial totals in Table~\ref{tab:timebudget}. Since benign inputs are rarely flagged, their cost remains close to the probe alone, whereas adversarial inputs, flagged in $90\%$ or more of cases, additionally incur the strong purification.

\begin{table}[t]
\centering
\caption{Average per-utterance running time (s).}
\label{tab:timebudget}
\small
\setlength{\tabcolsep}{5pt}
\begin{tabular}{lccc}
\toprule
& AASIST & RawGAT-ST & Res-TSSDNet \\
\midrule
Probe purification ($t_g$)  & 6.05  & 9.07  & 9.07  \\
Strong purification ($t_p$) & 42.34 & 45.36 & 24.19 \\
Scoring and transform       & 0.03  & 0.05  & 0.02  \\
Benign flag rate            & 5.1\%  & 2.1\%  & 23.9\% \\
Adversarial flag rate       & 93.0\% & 94.4\% & 90.0\% \\
\midrule
Benign total      & 8.23  & 10.06 & 14.87 \\
Adversarial total & 45.46 & 51.94 & 30.87 \\
\bottomrule
\end{tabular}
\end{table}

\section{Limitations}
\label{sec:limitations}

While our current study demonstrates that DGAP improves the robustness of audio deepfake detectors against diverse adversarial perturbations, it also has several limitations. First, under a defense-aware adaptive attack, the main failure mode is gate evasion, where the adversary can optimize the perturbation so that the score gap remains below the gating threshold and the strong purification is bypassed. Although this constraint still limits the attack compared with the undefended setting, future work should investigate stronger gating strategies, such as randomized probe levels, multiple probe strengths, or additional consistency checks. Second, diffusion-based purification introduces non-negligible computational cost. DGAP reduces this cost by applying strong purification only to flagged inputs, but every input still requires a light probe purification. More efficient samplers, distillation, or lightweight audio-specific diffusion backbones could make adaptive purification more practical for real-time deployment. Finally, our evaluation focuses on pretrained detectors and perturbation-based attacks. Future work can extend the framework to broader deepfake generation methods, unseen datasets, over-the-air perturbations, and joint attacks that combine synthesis and adversarial optimization.

\section{Conclusion}
We present DGAP, a detection-guided adaptive purification framework for robust audio deepfake detection under adversarial perturbations. Our method uses the detector's own response to decide when purification is needed, applying strong diffusion-based purification only to flagged inputs while leaving the pretrained detector unchanged and benign inputs nearly unaffected. Across three deepfake detectors and three adversarial attack settings, our framework achieves the strongest overall defense performance compared with existing defenses, while remaining effective under a defense-aware adaptive attack.

\section*{Acknowledgment}
This work was supported in part by the U.S. National Science Foundation grant CNS-2310207.

\bibliographystyle{IEEEtran}
\bibliography{references}
\end{document}